\documentclass{emulateapj}

\usepackage{psfig}
\usepackage{natbib}
\shorttitle{Fluorescent Molecular Hydrogen in IC 63}
\shortauthors{France et al.}

\begin{document}


\title{Fluorescent Molecular Hydrogen Emission in IC 63: 
$FUSE$, HUT, and Rocket Observations}


\author{K. France, B-G Andersson, S. R. McCandliss, and P. D. Feldman}
\affil{Department of Physics and Astronomy, Johns Hopkins University,
    Baltimore, MD 21218}





\begin{abstract}
We present far-ultraviolet observations of IC 63, an emission/reflection
nebula illuminated by the B0.5IV star $\gamma$~Cassiopeia, located 1.3~pc from
the nebula.  Molecular hydrogen
fluorescence was detected first in IC 63 by $IUE$ and 
later at shorter wavelengths by $ORFEUS$.
Here we present $Far$ $Ultraviolet$ $Spectroscopic$ $Explorer$ ($FUSE$) observations
towards three locations in the nebula, complemented by Hopkins Ultraviolet
Telescope (HUT) data on the central nebular position.  In addition, 
we present a sounding rocket calibration of a $FUSE$ spectrum of 
$\gamma$~Cas.  Molecular hydrogen fluorescence is
detected in all three FUSE pointings. The intensity of this emission as
well as the contributions from other species are seen to vary with position.
The absolute flux calibration of the sounding rocket data allows
us to reliably predict the radiation field incident on IC~63.  We use these
data to test models of the fluorescent process.  Our modeling resolves the
perceived discrepancy between the existing ultraviolet
observations and achieves a satisfactory agreement with the H$_{2}$ rotational
structure observed with $FUSE$.
\end{abstract}



\section{Introduction}

Ultraviolet fluorescence is the initial step in the 
process that gives rise to the near-infrared (IR) emission
spectrum of photo-excited molecular hydrogen (H$_{2}$) observed in a 
wide range of astronomical environments. 
Hydrogen molecules make the transition to an excited electronic state
(predominantly $B$$^{1}\Sigma^{+}_{u}$~and~$C$$^{1}\Pi_{u}$)
by absorbing ultraviolet (UV) photons.  The transition back to the 
ground electronic state produces the characteristic 
ultraviolet spectrum and leaves the molecules in excited rovibrational
levels.  The near-IR lines are emitted as the molecules return
to the ground vibrational level through quadrupole transitions. 
Infrared emission lines
can also be populated collisionally, and
are a common diagnostic used to probe the molecular gas
phase of many classes of 
astronomical objects (e.g., star forming regions: Luhman et al. 1994, 
Chrysostomou et al. 1993; reflection nebulae: Gatley et al. 1987, Takami et al.
2000; planetary nebulae: Zuckerman \& Gatley 1988, Luhman \& Rieke 1996).~\nocite{gatley87,takami00,luhman94,chrys93,zuckerman88,luhman96}  
Ultraviolet emission lines of H$_{2}$ cannot be produced thermally because 
the molecules would dissociate before the upper electronic states could
be populated,
so they must be pumped by ultraviolet photons or non-thermal electrons.
The detection of these lines is a clear indication of non-thermal
excitation occurring in at least some portion of the molecular gas.

The far-ultraviolet emission from molecular hydrogen was first predicted
to be detectable in diffuse objects by~\citet{duley80}.  \cite{witt89}
detected this emission in IC 63 with the Short Wavelength Primary (SWP) camera
on $IUE$, representing the first astronomical detection of the ultraviolet H$_{2}$
fluorescence spectrum pumped by a continuum 
source.  \citet{luhman97} reported the detection of the near-infrared emission
spectrum of fluorescent H$_{2}$, making IC 63 the first 
object seen to exhibit both the ultraviolet and infrared emission
from H$_{2}$
excited by ultraviolet continuum photons.  \citet{hurwitz98} 
presented the first spectrum of IC 63 at wavelengths shorter than the $IUE$ bandpass, using
the Berkeley Extreme and Far-ultraviolet Spectrograph aboard $ORFEUS-II$.

\begin{figure}
\begin{center}
\epsscale{1.2}
\rotatebox{90}{
\plotone{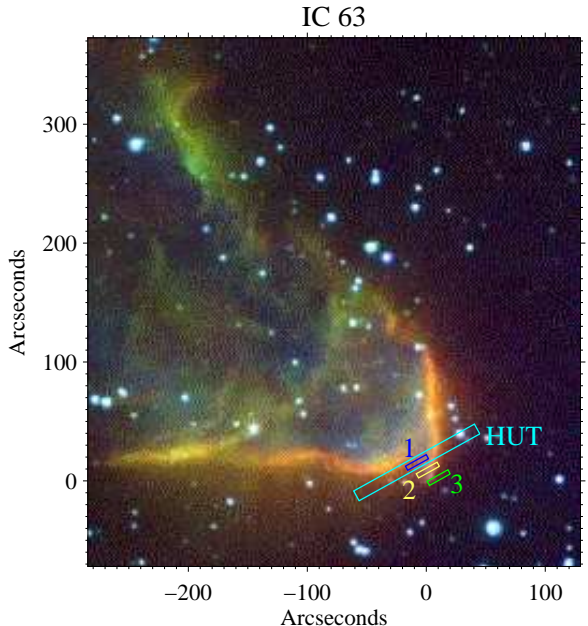} }
\caption{\label{slits} IC 63 with relevant aperture overlays.  $\gamma$~Cas
is located approximately 20\arcmin~to the southwest.  The image was
taken with the WIYN 0.9-meter, courtesy of Eric B. Burgh.}
\end{center}
\end{figure}

IC 63 is a bright emission/reflection nebula illuminated by the hot star
$\gamma$~Cas (HD 5394), a lightly reddened B0.5 IV
star
($E_{B~-~V}$~=~0.03; Hurwitz 1998).  Assuming that the star and the nebula
are co-spatial at a distance of $\approx$~200 pc, the bright optical
nebula is projected to be 1.3 pc from the star (Figure~\ref{slits}).   
A number of arguments are given in the literature
to support fluorescent pumping by the ultraviolet continuum
of $\gamma$~Cas as the process that gives rise to the H$_{2}$ 
emission observed by Witt et al. (1989), Luhman et al. (1997),
and Hurwitz (1998).~\nocite{witt89,luhman97,hurwitz98}  \citet{witt89} present 
calculations of the energy budgets for competing processes such as a
collisional excitation from a stellar wind or non-thermal excitation
from electrons produced by stellar X-rays, and find that these 
explanations are insufficient to reproduce the nebular brightness 
observed in IC 63.  The infrared H$_{2}$ line ratios seen by~\citet{luhman97}
are consistent with fluorescence rather than collisional excitation.
Hurwitz (1998) discounts the possibilities of shocks by the narrow
line widths seen in sub-mm molecular observations~\citep{jansen94}, 
but finds that models of UV fluorescence overpredict the
amount of far-UV emission observed.  
\citet{hurwitz98} found that the 
absolute flux of the far-UV emission lines were an order of magnitude
fainter than what would be expected based on the longer wavelength
ultraviolet lines.

We present the results of 
$Far$ $Ultraviolet$ $Spectroscopic$ $Explorer$ ($FUSE$) and 
Hopkins Ultraviolet Telescope (HUT) observations
of IC 63.  The high spectral resolution of $FUSE$ enables us to separate fluorescent emission
from the dust scattered continuum and resolve the lines into 
their individual rotational components.  These
results mark only the second object in which continuum-pumped fluorescence
has been resolved below 1150~\AA~(IC 405, France et al. 2004).~\nocite{france04} 
The HUT observation covers the entire 912~--~1650~\AA\ region
spanned by the fluorescent emission.  Additionally, 
we present the far-UV spectrum of the incident radiation field 
from $\gamma$~Cas obtained with a rocket-borne spectrograph.
This data has been used to calibrate a high-resolution scattered light spectrum 
from $FUSE$, allowing us to model the fluorescent H$_{2}$ using realistic
physical parameters for the exciting radiation field.

\section{Far-Ultraviolet Observations}

\subsection{$FUSE$ Observations}
\begin{figure*}
\begin{center}
\epsscale{1.2}
\rotatebox{90}{
\plotone{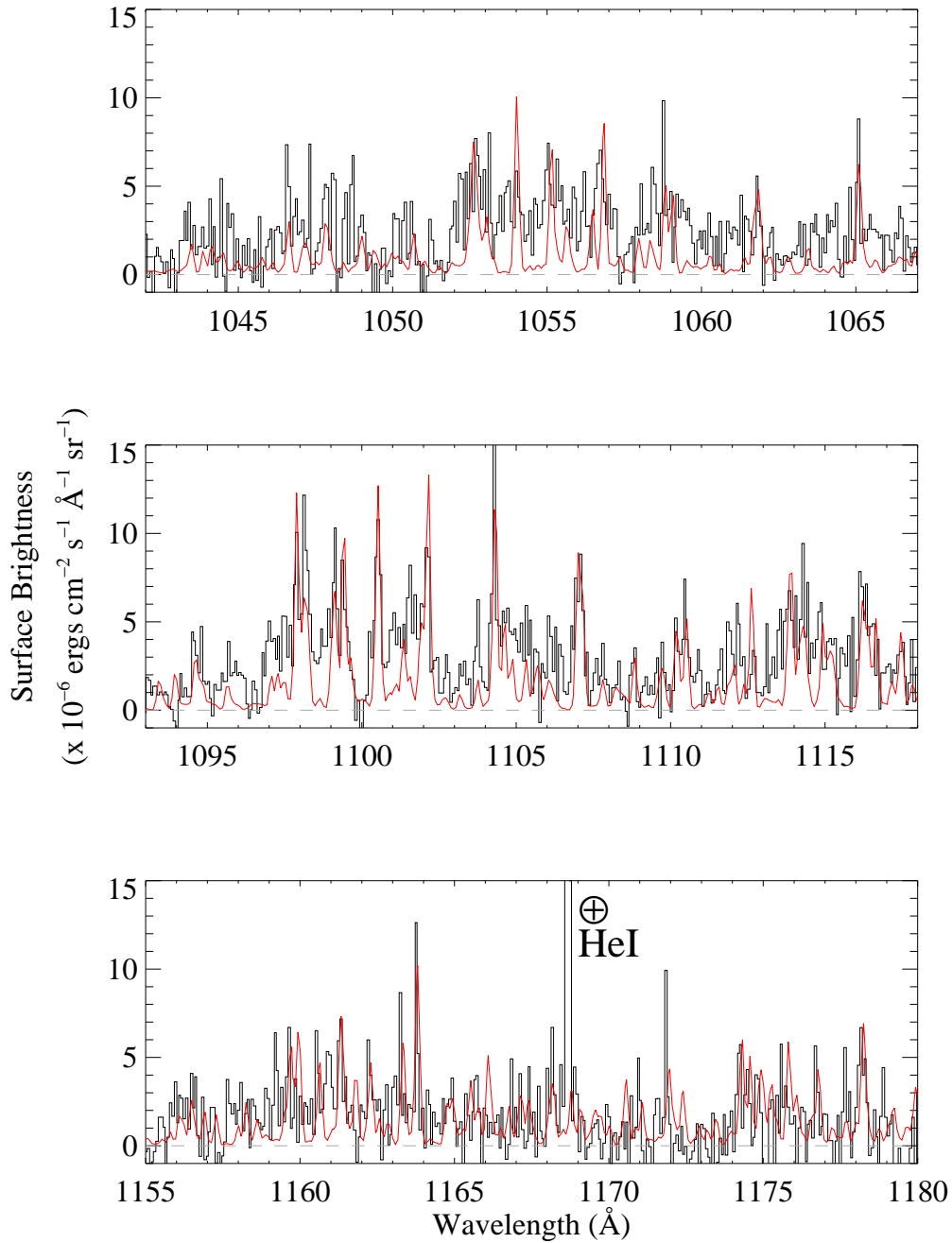} }
\caption{\label{fuseneb}  25~\AA\ windows of the $FUSE$ spectra of POS2 overplotted 
with the H$_{2}$ emission model described in \S 4.  The model spectrum
has been convolved to 0.15~\AA.  These spectra were acquired in the
LiF 1a and LiF 2a channels.}
\end{center}
\end{figure*}

IC 63 was observed by $FUSE$ on 2001 September 09 and 10 (see Moos et al. 2000 for a satellite description and Sahnow et al. 2000
for on-orbit performance characteristics).~\nocite{moos00,sanhow00} 
The (4\arcsec$\times$20\arcsec, filled-slit resolving power
$R$~$\sim$~8000) MDRS aperture was used 
to acquire spectra at three positions
in the nebula.
 POS1 was observed for 15.9 ks and
is located in the bright ``bullet-tip'' of IC 63. This position  
was chosen to overlap with previous pointings made with 
$IUE$, $ORFEUS$, and HUT
where ultraviolet fluorescence was detected
(Witt et al. 1989; Hurwitz 1998; this work).  POS2 is located along the 
limb of the bright optical emission, and was observed for 17.6 ks (Figure~\ref{fuseneb}).
POS3, observed for 31.5 ks, samples the region just outside the optical 
nebula.  Sample spectra for all three pointings are shown in Figure~\ref{fuselines}
and a summary of the $FUSE$ observations of IC 63 is given in Table 1.
The data were obtained in time-tagged (TTAG)
mode and have been reprocessed using the CalFUSE pipeline version 3.0.2.
In order to minimize channel drift in the LiF2 and the two SiC
channels, channel alignment (PEAK-UP) was performed once per 
orbit using the star HD~6417.
 $FUSE$ data acquired in TTAG mode
register the light through all three science apertures (LWRS, MDRS, and HIRS), 
and while no signal was detected through the HIRS slit, 
we obtained off-nebula spectra in the LWRS (30\arcsec$\times$30\arcsec) aperture.
It was necessary to redefine the regions used for background subtraction by the 
CalFUSE pipeline in order to accommodate the signal in the LWRS slit.
The nominal background regions used by 
CalFUSE include all regions on the detectors outside of the area of the 
spectrum from the primary slit.  We defined four smaller background regions 
on the detectors that avoided all three science apertures. 

\begin{deluxetable}{ccccc}
\tabletypesize{\small}
\tablecaption{Summary of $FUSE$ observations of IC 63. \label{fuse_sum}}
\tablewidth{0pt}
\tablehead{
\colhead{Position} & \colhead{Program}   & \colhead{RA (2000)}   &
\colhead{$\delta$ (2000)} & \colhead{Integration Time}  
\\ 
 &   &   \colhead{( $^{\mathrm h}$\, $^{\mathrm m}$\, $^{\mathrm s} )$} &
\colhead{(\arcdeg\, \arcmin\, \arcsec )}     & \colhead{(seconds)}  
}
\startdata
IC63-POS1 & B11201 & 00 59 01.29 & +60 53 17.9 & 15886  \\
IC63-POS2 & B11201 & 00 58 59.04 & +60 53 06.4 & 17564  \\
IC63-POS3 & B11203 & 00 58 56.79 & +60 52 54.9 & 31509  \\
 \enddata



\end{deluxetable}

$\gamma$~Cas exceeds the $FUSE$ bright target limit by almost two orders
of magnitude, but observing the instrumentally scattered spectrum of
the star allows the spectral characteristics to be recorded while protecting the 
primary science detectors.
$\gamma$~Cas was observed  
using this offset technique on 2002 December 16 under the $FUSE$ project
bright-object
test
program (S52107).  During this 3.1 ks observation, the LWRS aperture was
located $\sim$~1\arcmin~from $\gamma$~Cas (RA~=~00$^{h}$56$^{m}$50.65$^{s}$, 
$\delta$~=~+60$^{\circ}$42\arcmin53.0\arcsec, J2000).  Spectra were obtained on the
2A and 2B detector segments, providing wavelength coverage from
916.6~--~1181.9~\AA.  A flux calibration was performed using the 
sounding rocket spectrum of $\gamma$~Cas described below.

\subsection{Description of the HUT Observation}
IC 63 was observed by HUT, aboard $Astro$-1, on 1990 December 08.
1668 seconds of nighttime data were obtained in the 830-1880~\AA\
bandpass using the primary (9.4\arcsec$\times$116\arcsec) aperture
(spectrum ic63\_144, Figure~\ref{hutspec}).  A description of the HUT 
instrument and data reduction can be found in~\citet{hut1}.  
The spectrum was acquired from the Multi-Mission Archive at the 
Space Telescope Science Institute (MAST).
The pointing is 
identical to where \citet{witt89} detected ultraviolet H$_{2}$ fluorescence
which is coincident with the $FUSE$ POS1.

\subsection{Sounding Rocket Observation of $\gamma$~Cas}

$\gamma$~Cas was observed by a rocket-borne spectrograph in 
2003 December.  This was the first flight of the Long-Slit Imaging
Dual-Order Spectrograph (LIDOS, McCandliss et al. 2003).~\nocite{lidos}  
LIDOS employs two
complimentary ultraviolet-sensitive detectors to achieve a large dynamic
range in flux for imaging spectroscopy.  An updated version of the 
Faint Object Telescope (0.4-meter diameter, f/15.7 Dall-Kirkham;
Hartig et al. 1980; McCandliss et al. 1994; France et al. 2004)~\nocite{fot,srm94,france04} focuses light at the 
entrance aperture of the spectrograph, a mirrored slit-jaw into which a 
long-slit (10\arcsec~$\times$~300\arcsec) is etched.  The light entering the 
600 mm diameter Rowland spectrograph is dispersed by a silicon carbide grating.  
Far-UV light diffracted in the +1 order (900~--~1590~\AA) is 
recorded by a windowless $\delta$-doped CCD that is sensitive to bright objects
whereas the --1 order UV light (930~--~1680~\AA) is recorded by a 
photon-counting micro-channel plate (MCP) detector with a CsI photocathode, 
read out by a double delay-line anode~\citep{mcphate99}.  The MCP detector has a low 
background equivalent flux, extending our detection limit towards 
fainter diffuse objects.

LIDOS was launched aboard a Black Brant IX sounding rocket (NASA 36.208~UG)
from White Sands Missile Range, New Mexico, 
on 2003 December 16 at 20:00 MST.  The pointing was obtained 
using an axially mounted onboard startracker with a field of view of 
$\pm$~2$^{\circ}$.
Target acquisition was within a few arcminutes of the nominal pointing, 
and this field was relayed to the ground in real-time through a Xybion TV 
camera imaging the slitjaw (20\arcmin~field-of-view).  Fine adjustments 
(e.g. placing the star in the slit) were 
performed via commands to the ACS in real-time by a ground based
operator.  $\gamma$~Cas was placed in the 
spectrograph slit near T+200 seconds and a 28 second exposure was obtained
with the CCD, shown in Figure~\ref{rktstar}.  Bias and dark frames were also obtained in-flight.  Following
the CCD exposure, the MCP high-voltage was turned on and pointing offsets were
made to both IC 59 and IC 63.  The integration time was sufficiently short on 
the two nebulae that only a scattered light spectrum of $\gamma$~Cas
was detected at the position of IC 63.

The rocket data were analyzed using IDL code customized to read the data as
supplied by the telemetry system.  A background subtraction can be made by 
measuring the flux on the detector adjacent to where the star was located.
The data are then calibrated with 
measurements of the telescope mirror reflectivities and spectrograph
quantum efficiencies, measured both before and after flight in the calibration 
facilities located at The Johns Hopkins University.
The high levels of CCD background and fixed pattern noise, 
combined with a telescope focus problem experienced during flight, prevent a 
detailed analysis of the stellar spectrum, but the flux calibration can 
be transferred to $FUSE$ LWRS observations of the scattered 
spectrum of $\gamma$~Cas described above.  Spectral windows with appreciable 
signal-to-noise in the rocket data were transferred to the $FUSE$ spectrum, 
with the long wavelength end required to match the $IUE$ spectrum of $\gamma$~Cas.
This flux-calibrated spectrum of HD 5394, shown in Figure~\ref{fusestar}, is used as the input
spectrum for the molecular hydrogen fluorescence model described in Section 4.

\section{Nebular Line Identification}

\begin{figure}[b]
\begin{center}
\epsscale{0.6}
\rotatebox{90}{
\plotone{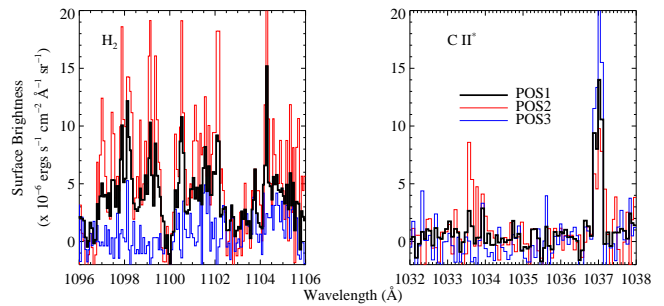} }
\caption{\label{fuselines}  Line strengths seen in the $FUSE$
spectra vary with position. These two regions are representative
of the trends followed by the molecular hydrogen and the ionic species.
H$_{2}$ shows a peak at POS2 (shown in red), falling to 
nearly zero at POS3 (shown in blue).  The ionic lines, represented 
here by \ion{C}{2}$^{*}$~$\lambda$1037, display the opposite behavior (with the 
exception of \ion{N}{2}$^{**}$~$\lambda$1085). For this comparison, the spectra
were shifted to a common background level.}
\end{center}
\end{figure}

Our $FUSE$ observations show a detection of fluorescent H$_{2}$ clearly at POS1 and POS2, and 
there is a tentative detection at POS3.  The strongest molecular hydrogen 
line complexes are centered on 1055, 1100, 1115, and 1161 \AA.  
\ion{C}{2}$^{*}$ $\lambda$1037 and \ion{N}{2}$^{*}$/\ion{N}{2}$^{**}$ $\lambda$1085 are seen strongly at 
all three positions.  \ion{S}{3}$^{**}$ $\lambda$1021 is seen seen clearly 
at POS1 and POS3.   
Interestingly, we do not detect the ground state transitions of 
\ion{C}{2} ($\lambda$1036) and \ion{S}{3} ($\lambda$1012), and only marginally detect 
\ion{N}{2}~$\lambda$~1084 at POS1 and POS3.   \ion{C}{2} $\lambda$1036, for example, would be
expected to be roughly half as strong as the excited \ion{C}{2}$^{*}$
line (shown in Figure~\ref{fuselines}), detectable at the S/N of these observations.  We attribute
this behavior to self-absorption by ground state ions within 
the nebula.  This is supported by the off-nebula spectra simultaneously
acquired in the LWRS aperture.  While 
\ion{C}{2} $\lambda$1036 is still undetected, the lower lying transitions
of \ion{N}{2} ($\lambda$1084) and \ion{S}{3} ($\lambda$1012 and 
$\lambda$1015) are seen.  This suggests that the ionized nebula extends
beyond both the bright optical emission and the molecular emission, 
traced by the H$_{2}$ fluorescence and the CO maps of~\citet{jansen94}.
This picture agrees with the findings of~\citet{karr05}, 
who find an ionization front in the direction of $\gamma$~Cas
giving way to molecular material (traced by PAH emission) deeper into the nebula.

Line strengths are measured by integrating over wavelength
and subtracting the integrated continuum of a nearby region
to 
account for the large uncertainty in the image background subtraction.
The ionic and the H$_{2}$ line strengths, listed in Table 2, display inverse relationships 
with distance from the $IUE$/HUT position.  The ions show a dip in strength 
at POS2, rising  to a maximum at POS3 whereas the molecular hydrogen lines
show a peak at POS2, and fall to nearly zero beyond the optical nebula (POS3).
The exception to this anti-correlation is \ion{N}{2}$^{**}$ $\lambda$1085, 
which roughly follows the molecular trend, but does not 
truncate as sharply as the H$_{2}$ at POS3.  
Errors are difficult to estimate due to the differences in the size and
location of the extraction regions used, we conservatively assign errors of
$\pm$~25\% to the quoted values.
We measure the widths of the ionic lines to be $\sim$~0.1~--~0.15~\AA, 
and find them to be at their rest wavelengths.  This is consistent
with the 1~--~2~km~s$^{-1}$ velocities found by~\citet{karr05}.
We would not expect to 
measure a wavelength shift at the resolution of the $FUSE$ instrument.
The molecular hydrogen lines are found to be somewhat broader, as discussed in Section 5.
Finally, we have found an 
unidentified triplet feature near 962 \AA\ whose
brightness increases in the MDRS aperture from POS1 to POS3.  
This feature is most likely \ion{P}{2} $\lambda$961.04/962.12/962.57, however it does not exhibit
the spatial variation seen in any of the other species.
This feature is not seen in the LWRS aperture, confining its spatial extent
to within~$\approx$~200\arcsec~of the optical nebula.

\begin{deluxetable}{ccccc}
\tabletypesize{\small}
\tablecaption{Line strengths seen in the $FUSE$ spectra 
	(in units of ergs s$^{-1}$ cm$^{-2}$ sr$^{-1}$). 
	The displacements are realtive to the HUT pointing in IC 63.
	Errors are estimated to be $\pm$~25\%.\label{fuse_lines}}
\tablewidth{0pt}
\tablehead{
\colhead{Line} & \colhead{$\lambda$ (\AA)}   &    &
\colhead{Line Strengths} &   \\ 
 &   &   \colhead{POS1 (0\arcsec)} &
\colhead{POS2 (28\arcsec)}     & \colhead{POS3(56\arcsec) }
}
\startdata
H$_{2}$  & 1055 & 5.2 x 10$^{-6}$ & 9.5 x 10$^{-6}$ & $\sim$~0  \\
H$_{2}$  & 1100 & 3.1 x 10$^{-5}$ & 6.3 x 10$^{-5}$ & 5.9 x 10$^{-6}$  \\
\ion{C}{2}$^{*}$  & 1037 & 2.8 x 10$^{-6}$ & 1.9 x 10$^{-6}$ & 4.0 x 10$^{-6}$  \\
\ion{N}{2}  & 1084 & 1.8 x 10$^{-6}$ & 3.7 x 10$^{-7}$ & 7.3 x 10$^{-7}$  \\
\ion{N}{2}$^{*}$  & 1085 & 4.6 x 10$^{-6}$ & 4.0 x 10$^{-6}$ & 4.1 x 10$^{-6}$  \\
\ion{N}{2}$^{**}$  & 1085 & 9.5 x 10$^{-6}$ & 1.0 x 10$^{-5}$ & 5.0 x 10$^{-6}$  \\
\ion{P}{2}\tablenotemark{a}  & 962 & 4.8 x 10$^{-6}$ & 1.2 x 10$^{-5}$ & 2.5 x 10$^{-5}$  \\
\ion{S}{3}$^{**}$  & 1021 & 1.5 x 10$^{-6}$ & 8.5 x 10$^{-7}$ & 1.9 x 10$^{-6}$  \\
\enddata
\tablenotetext{a}{Tentative identification}

\end{deluxetable}

\begin{figure}
\begin{center}
\epsscale{0.7}
\rotatebox{90}{
\plotone{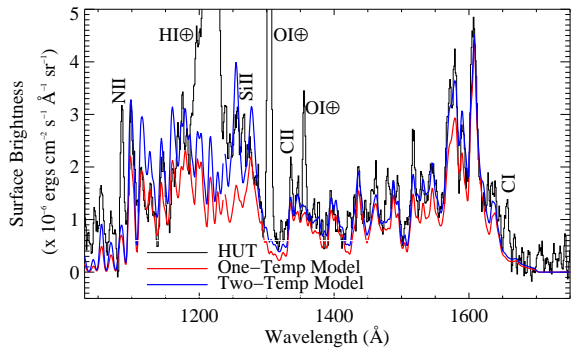} }
\caption{\label{hutspec}  The HUT spectrum of IC 63 following a  
scattered light subtraction of the form $aF_{\star}\lambda^{\beta}$, 
as described in the text.  The molecular hydrogen 
fluorescence models described in the text have been convolved to a 
resolution of 4~\AA.  A single temperature model (620~K) is shown in red to 
demonstrate the need for two temperature components (620~+~2500~K), plotted in blue.
Nebular
atomic emission lines are labeled.}
\end{center}
\end{figure}

\begin{figure}[b]
\begin{center}
\epsscale{0.6}
\rotatebox{90}{
\plotone{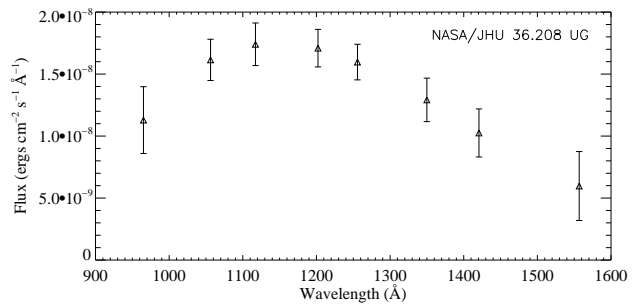} }
\caption{\label{rktstar} Low resolution rocket-borne CCD spectrum of $\gamma$~Cas.
The absolute calibration of this spectrum was used to transfer calibration
to the scattered light spectrum obtained by $FUSE$.}
\end{center}
\end{figure}

The strongest nebular feature seen in the
HUT
spectrum is the $\lambda\lambda$1578/1608 H$_{2}$ emission.  The 
short-wavelength hydrogen bands are suggested, but are hard to identify 
unambiguously at the HUT resolution ($\geq$~3~\AA, depending on the 
filling fraction within the slit).  \ion{C}{2}~$\lambda$1335 and
\ion{N}{2}~$\lambda$1085 are seen as well a scattered light
contribution from nebular dust.  The HUT spectrum gives us the opportunity
to compare the emission from H$_{2}$ both above and below Ly-$\alpha$, 
necessary for a thorough test of molecular hydrogen fluorescence
models
(see \S~4).  The HUT spectrum of IC 63 is shown in Figure~\ref{hutspec}
following a scattered light subtraction similar to the correction
described by~\citet{witt89}.  The scattered light was assumed to have a 
form $aF_{\star}\lambda^{\beta}$ where $F_{\star}$ was made from a composite
spectrum of $\gamma$~Cas created from archival $IUE$ data and the $FUSE$
spectrum described above.  The $FUSE$ data were degraded to the $IUE$
resolution and then joined to created a composite stellar spectrum spanning 
the entire range ($\approx$~900~--~1900~\AA) covered by HUT.
Following this procedure, we find that $a$~=~1.3~$\times$~10$^{-5}$ 
and $\beta$~=~1.5.  

\section{Fluorescent Molecular Hydrogen Model} \label{model}
Synthetic spectra of fluorescent emission from molecular hydrogen can 
be made by computing the radiative excitation rates into the
excited electronic states of H$_{2}$.  
Such models assume a ground electronic
state population, then use photoexcitation cross sections and 
an incident radiation field to calculate the rovibrational levels of 
the upper electronic state 
(predominantly $B$$^{1}\Sigma^{+}_{u}$~and~$C$$^{1}\Pi_{u}$).
The molecules will then return to the ground electronic state following
the appropriate selection rules and branching ratios, producing
the observed ultraviolet emission lines and leaving the molecules
in excited rovibrational levels.  \citet{sternberg} has described
calculations of the far-ultraviolet spectrum of H$_{2}$. However,~\citet{hurwitz98}
finds that these models overpredict the observed short-wavelength intensity
by roughly an order of magnitude.  Such trends are hinted at in the 
model spectrum of~\citet{witt89}, although it seems that their $IUE$
observations
did not go deep enough into the far-ultraviolet to see this effect fully.

We have adopted a modified version of the synthetic
molecular hydrogen emission spectrum presented by~\citet{wolven97} to model fluorescence induced by 
solar Ly-$\alpha$ at the Shoemaker-Levy 9 impact site on Jupiter.  These 
models include photoexcitation cross-sections computed using the line transition
probabilities from~\citet{abgrall93a,abgrall93b}.  
We adopted a Doppler $b$-value of 2 km s$^{-1}$.  
Sub-mm molecular line widths are seen to be quite narrow~\citep{jansen94,hurwitz98}, 
and 2 km s$^{-1}$ is the smallest $b$-value that is compatible with
the $H_{2}ools$ optical depth templates described by~\citet{h2ools}.
The model is largely insensitive to the actual choice of $b$-value
for values of a few km s$^{-1}$.
The ratio of atomic to molecular
hydrogen column densities is fixed to be 0.1, extrapolating the values
found in the translucent cloud survey of~\citet{rachford02}.
Additionally, the
Wolven models allow for absorption out of upper vibrational states ($\nu~\geq$~0)
and include a first-order correction for self-absorption by H$_{2}$
at wavelengths shorter than 1100~\AA.  
These models also take into account electronic transitions
to the $B^{'}$, $B^{''}$, $D$, and $D^{'}$ states, although their
relative contribution to the resultant spectrum is small.  Finally, 
transitions to predissociating states and vibrational states that result in dissociation
($\nu^{''}~>$~14, the vibrational continuum) are considered.  
For the model developed for IC 63, we find a dissociation fraction
of 16.8\%.

\begin{figure}[b]
\begin{center}
\epsscale{0.6}
\rotatebox{90}{
\plotone{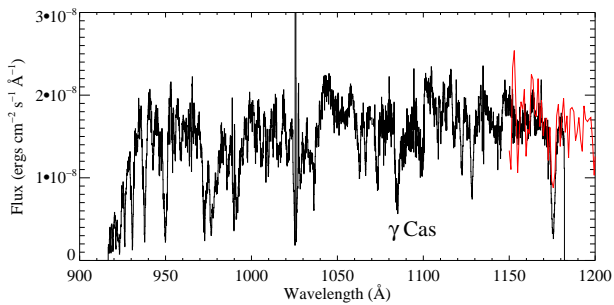} }
\caption{\label{fusestar}The $FUSE$ scattered light spectrum of $\gamma$~Cas (HD 5394) was 
flux calibrated by comparison with a low resolution rocket-borne spectrograph.
The $IUE$ spectrum of $\gamma$~Cas is shown in red in the overlap region.
The calibrated $FUSE$ spectrum was used as the incident radiation field in the 
fluorescence model described in the text.}
\end{center}
\end{figure}

There are several important differences between
the model described in~\citet{wolven97} and the one presented here.  The first
is that the we only consider the photo-induced fluorescence, no electron-impact
induced contribution is included.  \citet{witt89} has
determined that neither high-energy electrons nor a stellar wind
make a large contribution to the H$_{2}$ excitation. \citet{hurwitz98}
has provided additional evidence against contributions from
high-energy electrons.
The solar Ly-$\alpha$ profile
that was used as the excitation spectrum in~\citet{wolven97} has been replaced by the flux calibrated
$FUSE$ spectrum of $\gamma$~Cas, described in Section~2.1 and shown in Figure~\ref{fusestar}.
We only include the 917~--~1182~\AA\ region covered, but with the majority of
the upper states populated through absorption between the Lyman limit and
the (0~-~0) band near 1108~\AA, we expect errors induced by this approximation to be minor.

The fluorescence code described here uses two temperature components, the
rotational temperature quoted in~\cite{habart04}, 620~K, and a higher ``non-thermal''
vibrational temperature of 2500~K that was required to find agreement with the HUT
data.  The feature near 1578~\AA\ that is produced in transitions to the 
vibrational continuum of the ground electronic state was underpredicted 
without a  high-temperature component (Figure ~\ref{hutspec}).  

The physical basis for this vibrational temperature 
component is to mimic the effects of multiple pumping by UV photons.
As described above, the fluorescence process can leave the molecule in 
a number of excited vibrational levels of the ground electronic state.
Since the quadrupole decay rates are comparatively slow, the molecule
can absorb another UV photon before it has returned to the ground vibrational state.
We simulate this excited vibrational population with the second, higher
temperature component.
The 2500~K temperature applies only to the distribution of vibrational
levels and is not coupled to the population of the rotational states.
This decoupling occurs because of the selection rules for the electronic transitions
of H$_{2}$.  The molecule can only change its rotational level ($\Delta$$j$)
by -1, 0, or +1, where as the vibrational levels can change ($\Delta$$v$) by any 
value during the electronic transitions.

We also use the column density from 
Habart et al. (2004; 5~$\times$~10$^{21}$ cm$^{-2}$) 
in the fluorescence code.  We found that 
the model required an additional molecular hydrogen absorption component
(of the same column and $b$-value used in the emission model)
to agree with the HUT data at the 1100~\AA\ H$_{2}$ emission band.
This additional absorption component was calculated using the 
$H_{2}ools$ templates~\citep{h2ools}.  The $FUSE$ spectra are better fit without
this added absorption.  As the bulk of the H$_{2}$ emission seen in the HUT
data is unaffected by this additional absorption component, it is 
difficult to assign the a level of significance to this result, especially
given the uncertainties discussed below. 
Assuming that the interstellar extinction
along the line of sight to IC 63 is small, as is the case for $\gamma$~Cas, 
we do not include a reddening correction.

\section{Discussion}

\subsection{Spatial Extent of the Fluorescing Region} 
        We can gain an estimate of the size of the region from which the H$_{2}$ 
fluorescence emanates by comparing the line fluxes observed by $FUSE$, $ORFEUS$ and
HUT.  The slit areas of these three instruments were 80, 530 \& 1090 \arcsec$^2$, 
respectively.  Based on the MAST archival spectra we find that the integrated 
fluxes for the line complexes at $\sim$1055\AA\ and $\sim$1100\AA\ we find the 
the fluxes observed by $ORFEUS$ and HUT are consistent with each other at $\sim$2 
and 3$\times$10$^{-13}$ ergs s$^{-1}$ cm$^{-2}$, respectively.  The same two regions 
yield integrated fluxes of $\sim$5 and 7$\times$10$^{-14}$ ergs s$^{-1}$ cm$^{-2}$ in the 
$FUSE$ observations. Although such a comparison is complicated by the fact that 
the slit geometries differ, with $FUSE$ and HUT having rectangular slits while 
$ORFEUS$ employed a circular slit, we find that the 
region giving rise to the H$_{2}$ fluorescence extends beyond the area of the
$FUSE$ MDRS slit, but does not fill the HUT or $ORFEUS$ slits.  If we assume that
the emitting region is of uniform surface brightness, we can use the flux
ratios to estimate that the $ORFEUS$ slit was about 60 \% filled.  As the emission 
seen by $FUSE$ towards POS2 is of similar strength as POS1 we 
conclude that the center of the emitting region is probably located between 
these two positions.  If we assume that the emitting region is a circle, 
centered between our POS1 and 2, we can then estimate that the 
diameter of the emitting region is about 30''.  

\subsection{H$_{2}$ Modeling Results}

Figures 2 and 4 show a comparison of the nebular H$_{2}$ emission 
spectrum with the model described above.  The HUT data allow us
to test this model both above and below Lyman-$\alpha$ for the 
first time while the  $FUSE$ spectrum gives us the opportunity to 
study the short wavelength H$_{2}$ spectrum in detail.  In each instance, the
model requires an offset in the absolute flux to agree with 
the observations.  Our H$_{2}$ model is seen to roughly 
fit the spectrum of IC 63 both below Ly-$\alpha$ and at the longest
ultraviolet wavelengths ($\sim$~1600~\AA), resolving the
order of magnitude discrepancy
between the relative strengths of the ultraviolet emission components
seen in~\cite{hurwitz98} and suggested in the spectrum presented by~\cite{witt89}.
Our model still shows some deviation at short wavelengths, but the
differences are a factor of two at the most.  H$_{2}$ emission in the 
1115 and 1143~\AA\ bands are better fit with the 
single temperature model, shown in red in Figure~\ref{hutspec}, despite the 
underprediction of this model elsewhere in the spectrum.
Disagreement between the 
model and the data could be due to an 
imperfect scattered light subtraction and/or an unknown amount of interstellar 
extinction.  To achieve the fit to the HUT spectrum, 
the model requires an overall scale  factor of 11, yet his is probably
an under representation of the actual scaling required because the
molecular emission is unlikely to be filling the entire HUT aperture~(\S~5.1). 

We find that the model shows relatively good agreement with the 
strengths of the rotationally resolved lines in the $FUSE$ data.  The
lines appear broader than would be expected from a purely instrumental
effect.  We expect a filled-slit resolution of $\approx$~0.125~\AA\ 
in the MDRS aperture, yet some H$_{2}$ lines are broader than 0.25~\AA.
We tentatively explain this as a redistribution
effect caused by the finite width of the absorbing transition~\citep{hummer62}.  The 
exact cause is beyond the scope of this paper and will be addressed in 
future work.  In contrast to the HUT spectrum, where the model needed
to be scaled up to fit the data, we find that in order to achieve a 
good fit to the H$_{2}$ lines observed in the $FUSE$ spectra, the model needs to 
be scaled down by a factor of 0.5.  

One possible explanation for the variations we find in the 
scaling needed to reach agreement between the model and the different
data sets could be the clumpy nature of the molecular gas in IC 63.
Recent work has found that molecular gas and dust tend to form 
dense knots in the presence of an intense ultraviolet radiation field~\citep{odell00,huggins02,france04}.
If we assume that such knots exist in IC 63, then the 30\arcsec~emitting 
region may be composed of dense clumps instead of 
smoothly distributed molecular gas.  This could reconcile the scaling
we require to bring our model into agreement with the data.  
The larger 
HUT aperture may have included several of these dense clumps, where as the 
smaller $FUSE$ MDRS aperture may have picked up fewer, or simply missed the 
brightest of these knots.  Differing contributions from dense regions in
IC 63 could explain the different scale factors needed to predict the
absolute flux seen in the nebular spectra.

\section{Summary}

We have presented far-ultraviolet spectroscopy of the emission/reflection
nebula IC 63.  The MDRS aperture on $FUSE$ was used to obtain high-resolution
spectra of three positions within the nebula across the 912~--~1187~\AA\
bandpass.  These data were complemented by an archival HUT spectrum
of IC 63 at lower resolution, extending the wavelength coverage to 
beyond 1700~\AA.  These data confirm the presence of a population
of fluorescing H$_{2}$ seen in previous studies.  Models of this emission
have been shown to overpredict the relative strength of the shortest
wavelength lines by as much as an order of magnitude.  We use these
data to develop a model that not only accurately describes the detailed
rotational structure of the emission lines, but resolves the 
perceived discrepancy between the shortest and longest ultraviolet 
wavelengths spanned by H$_{2}$.  This model incorporates a realistic
incident radiation field by using sounding rocket observations to 
calibrate a $FUSE$ scattered light spectrum of the exciting star, $\gamma$ Cas.
Our model finds satisfactory agreement with the spectral 
structure seen with $FUSE$ and finds relative consistency at all wavelengths.
Our model still does not predict the correct absolute flux that is observed, 
and the emission lines in the $FUSE$ spectra are broader than what
would be expected from instrumental effects alone.  We are addressing these
problems by developing a more sophisticated H$_{2}$ model that
takes the finite width of the absorbing transition into account.

\acknowledgments
We would like to thank Eric Burgh for the optical image of 
IC 63 and constructive comments.   It is a pleasure to acknowledge helpful discussions with Saavik Ford, Amiel
Sternberg and David Neufeld.
We appreciate the effort undertaken
by the $FUSE$ mission planning staff, particularly Martin England, to 
make the challenging observations of IC 63. We thank Russ Pelton for his dedication
to all aspects of the sounding rocket flight.  
We also wish to thank Derek Hammer for technical assistance.
We acknowledge 
the support of the NSROC team from Wallops Flight Facility and the Physical Science
Lab operated by New Mexico State University. 
$FUSE$ data were obtained under the Guest Investigator Program 
(NASA grant NAG510380) by the 
NASA-CNES-CSA $FUSE$ mission, operated by the Johns Hopkins University.
The HUT spectrum presented in this paper was obtained from the 
MultiMission Archive at the Space Telescope Science Institute.
The rocket spectrum of
$\gamma$~Cas was supported by 
NASA grant NAG5-5122 to the Johns Hopkins University.


\bibliography{ms}

\begin{thebibliography}{30}
\expandafter\ifx\csname natexlab\endcsname\relax\def\natexlab#1{#1}\fi

\bibitem[{{Abgrall} {et~al.}(1993{\natexlab{a}}){Abgrall}, {Roueff}, {Launay},
  {Roncin}, \& {Subtil}}]{abgrall93a}
{Abgrall}, H., {Roueff}, E., {Launay}, F., {Roncin}, J.~Y., \& {Subtil}, J.~L.
  1993{\natexlab{a}}, \aaps, 101, 273

\bibitem[{{Abgrall} {et~al.}(1993{\natexlab{b}}){Abgrall}, {Roueff}, {Launay},
  {Roncin}, \& {Subtil}}]{abgrall93b}
---. 1993{\natexlab{b}}, \aaps, 101, 323

\bibitem[{{Chrysostomou} {et~al.}(1993){Chrysostomou}, {Brand}, {Burton}, \&
  {Moorhouse}}]{chrys93}
{Chrysostomou}, A., {Brand}, P.~W.~J.~L., {Burton}, M.~G., \& {Moorhouse}, A.
  1993, \mnras, 265, 329

\bibitem[{{Davidsen} {et~al.}(1992){Davidsen}, {Long}, {Durrance}, {Blair},
  {Bowers}, {Conard}, {Feldman}, {Ferguson}, {Fountain}, {Kimble}, {Kriss},
  {Moos}, \& {Potocki}}]{hut1}
{Davidsen}, A.~F., {Long}, K.~S., {Durrance}, S.~T., {Blair}, W.~P., {Bowers},
  C.~W., {Conard}, S.~J., {Feldman}, P.~D., {Ferguson}, H.~C., {Fountain},
  G.~H., {Kimble}, R.~A., {Kriss}, G.~A., {Moos}, H.~W., \& {Potocki}, K.~A.
  1992, \apj, 392, 264

\bibitem[{{Duley} \& {Williams}(1980)}]{duley80}
{Duley}, W.~W. \& {Williams}, D.~A. 1980, \apjl, 242, L179

\bibitem[{{France} {et~al.}(2004){France}, {McCandliss}, {Burgh}, \&
  {Feldman}}]{france04}
{France}, K., {McCandliss}, S.~R., {Burgh}, E.~B., \& {Feldman}, P.~D. 2004,
  \apj, 616, 257

\bibitem[{{Gatley} {et~al.}(1987){Gatley}, {Hasegawa}, {Suzuki}, {Garden},
  {Brand}, {Lightfoot}, {Glencross}, {Okuda}, \& {Nagata}}]{gatley87}
{Gatley}, I., {Hasegawa}, T., {Suzuki}, H., {Garden}, R., {Brand}, P.,
  {Lightfoot}, J., {Glencross}, W., {Okuda}, H., \& {Nagata}, T. 1987, \apjl,
  318, L73

\bibitem[{{Habart} {et~al.}(2004){Habart}, {Boulanger}, {Verstraete},
  {Walmsley}, \& {Pineau des For{\^ e}ts}}]{habart04}
{Habart}, E., {Boulanger}, F., {Verstraete}, L., {Walmsley}, C.~M., \& {Pineau
  des For{\^ e}ts}, G. 2004, \aap, 414, 531

\bibitem[{{Hartig} {et~al.}(1980){Hartig}, {Fastie}, \& {Davidsen}}]{fot}
{Hartig}, G.~F., {Fastie}, W.~G., \& {Davidsen}, A.~F. 1980, \ao, 19, 729

\bibitem[{{Huggins} {et~al.}(2002){Huggins}, {Forveille}, {Bachiller}, {Cox},
  {Ageorges}, \& {Walsh}}]{huggins02}
{Huggins}, P.~J., {Forveille}, T., {Bachiller}, R., {Cox}, P., {Ageorges}, N.,
  \& {Walsh}, J.~R. 2002, \apjl, 573, L55

\bibitem[{{Hummer}(1962)}]{hummer62}
{Hummer}, D.~G. 1962, \mnras, 125, 21

\bibitem[{{Hurwitz}(1998)}]{hurwitz98}
{Hurwitz}, M. 1998, \apjl, 500, L67+

\bibitem[{{Jansen} {et~al.}(1994){Jansen}, {van Dishoeck}, \&
  {Black}}]{jansen94}
{Jansen}, D.~J., {van Dishoeck}, E.~F., \& {Black}, J.~H. 1994, \aap, 282, 605

\bibitem[{{Karr} {et~al.}(2005){Karr}, {Noriega-Crespo}, \& {Martin}}]{karr05}
{Karr}, J.~L., {Noriega-Crespo}, A., \& {Martin}, P.~G. 2005, \aj, 129, 954

\bibitem[{{Luhman} \& {Rieke}(1996)}]{luhman96}
{Luhman}, K.~L. \& {Rieke}, G.~H. 1996, \apj, 461, 298

\bibitem[{{Luhman} {et~al.}(1994){Luhman}, {Jaffe}, {Keller}, \&
  {Pak}}]{luhman94}
{Luhman}, M.~L., {Jaffe}, D.~T., {Keller}, L.~D., \& {Pak}, S. 1994, \apjl,
  436, L185

\bibitem[{{Luhman} {et~al.}(1997){Luhman}, {Luhman}, {Benedict}, {Jaffe}, \&
  {Fischer}}]{luhman97}
{Luhman}, M.~L., {Luhman}, K.~L., {Benedict}, T., {Jaffe}, D.~T., \& {Fischer},
  J. 1997, ApJL, 480, L133

\bibitem[{{McCandliss}(2003)}]{h2ools}
{McCandliss}, S.~R. 2003, \pasp, 115, 651

\bibitem[{{McCandliss} {et~al.}(2003){McCandliss}, {France}, {Feldman}, \&
  {Pelton}}]{lidos}
{McCandliss}, S.~R., {France}, K., {Feldman}, P.~D., \& {Pelton}, R. 2003, in
  Future EUV/UV and Visible Space Astrophysics Missions and Instrumentation.
  Edited by J. Chris Blades, Oswald H. W. Siegmund. Proceedings of the SPIE,
  Volume 4854, pp. 385-396 (2003)., 385--396

\bibitem[{{McCandliss} {et~al.}(1994){McCandliss}, Martinez, {Feldman},
  {Pelton}, {Keski-Kuha}, \& {Gum}}]{srm94}
{McCandliss}, S.~R., Martinez, M.~E., {Feldman}, P.~D., {Pelton}, R.,
  {Keski-Kuha}, R.~A., \& {Gum}, J.~S. 1994, in Proceedings of the SPIE, Vol.
  2011

\bibitem[{{McPhate} {et~al.}(1999){McPhate}, {Feldman}, {McCandliss}, \&
  {Burgh}}]{mcphate99}
{McPhate}, J.~B., {Feldman}, P.~D., {McCandliss}, S.~R., \& {Burgh}, E.~B.
  1999, \apj, 521, 920

\bibitem[{{Moos}(2000)}]{moos00}
{Moos}, H.~W. et~al. 2000, \apjl, 538, L1

\bibitem[{{O'Dell}(2000)}]{odell00}
{O'Dell}, C.~R. 2000, \aj, 119, 2311

\bibitem[{{Rachford} {et~al.}(2002){Rachford}, {Snow}, {Tumlinson}, {Shull},
  {Blair}, {Ferlet}, {Friedman}, {Gry}, {Jenkins}, {Morton}, {Savage},
  {Sonnentrucker}, {Vidal-Madjar}, {Welty}, \& {York}}]{rachford02}
{Rachford}, B.~L., {Snow}, T.~P., {Tumlinson}, J., {Shull}, J.~M., {Blair},
  W.~P., {Ferlet}, R., {Friedman}, S.~D., {Gry}, C., {Jenkins}, E.~B.,
  {Morton}, D.~C., {Savage}, B.~D., {Sonnentrucker}, P., {Vidal-Madjar}, A.,
  {Welty}, D.~E., \& {York}, D.~G. 2002, \apj, 577, 221

\bibitem[{{Sahnow}(2000)}]{sanhow00}
{Sahnow}, D.~J. et~al. 2000, \apjl, 538, L7

\bibitem[{{Sternberg}(1989)}]{sternberg}
{Sternberg}, A. 1989, ApJ, 347, 863

\bibitem[{{Takami} {et~al.}(2000){Takami}, {Usuda}, {Sugai}, {Kawabata},
  {Suto}, \& {Tanaka}}]{takami00}
{Takami}, M., {Usuda}, T., {Sugai}, H., {Kawabata}, H., {Suto}, H., \&
  {Tanaka}, M. 2000, ApJ, 529, 268

\bibitem[{{Witt} {et~al.}(1989){Witt}, {Stecher}, \& {Boroson}}]{witt89}
{Witt}, A.~N., {Stecher}, T.~P., \& {Boroson}, T. A. an d~{Bohlin}, R.~C. 1989,
  ApJL, 336, L21

\bibitem[{{Wolven} {et~al.}(1997){Wolven}, {Feldman}, {Strobel}, \&
  {McGrath}}]{wolven97}
{Wolven}, B.~C., {Feldman}, P.~D., {Strobel}, D.~F., \& {McGrath}, M.~A. 1997,
  \apj, 475, 835

\bibitem[{{Zuckerman} \& {Gatley}(1988)}]{zuckerman88}
{Zuckerman}, B. \& {Gatley}, I. 1988, \apj, 324, 501

\end{thebibliography}

\end{document}